\documentclass[11pt,a4paper,english]{article}
 \usepackage[T1]{fontenc}
 \usepackage[margin=2cm]{geometry}
 \usepackage[hidelinks]{hyperref}

\usepackage{graphicx}%
\usepackage{multirow}%
\usepackage{amsmath,amssymb,amsfonts}%
\usepackage{amsthm}%
\usepackage{mathrsfs}%
\usepackage[title]{appendix}%
\usepackage{xcolor}%
\usepackage{textcomp}%
\usepackage{manyfoot}%
\usepackage{enumitem}%
\usepackage{booktabs}%
\usepackage{algorithm}%
\usepackage{algorithmicx}%
\usepackage{algpseudocode}%
\usepackage{listings}%
\usepackage{tikz}
\usetikzlibrary{shapes,shadows,arrows}
\usepackage[frozencache,cachedir=.]{minted}
\usepackage{thm-restate}
\usepackage{cleveref}
\crefname{theorem}{Theorem}{Theorems}
\crefname{observation}{Observation}{Observations}
\crefname{lemma}{Lemma}{Lemmas}
\crefname{corollary}{Corollary}{Corollaries}
\crefname{proposition}{Proposition}{Propositions}
\crefname{example}{Example}{Examples}
\crefname{claim}{Claim}{Claims}
\crefname{table}{Table}{Tables}
\crefname{equation}{Inequality}{Inequalities}
\newtheorem*{remark*}{Remark}
\crefname{reductionrule}{Reduction rule}{Reduction rules}
\crefname{section}{Section}{Sections}
\crefname{figure}{Figure}{Figures}
\crefname{chapter}{Chapter}{Chapters}

\newtheorem{theorem}{Theorem}
\newtheorem{proposition}[theorem]{Proposition}%
\newtheorem{lemma}{Lemma}
\newtheorem{corollary}{Corollary}
\newtheorem*{claim*}{Claim}
\newtheorem*{definition*}{Definition}

\newcommand{\match}{\mathsf{Match}}
\newcommand{\UM}{\mathsf{UM}}

\newcommand{\FOA}{\mathsf{FOA}}
\newcommand{\FOAa}{\mathsf{FOA_{aux}}}
\newcommand{\FOB}{\mathsf{FOB}}

\newcommand{\Fair}{\mathsf{Fair}}
\newcommand{\MM}{\mathsf{MM}}
\newcommand{\ltp}{\mathsf{Last\_Transaction\_Price}}
\newcommand{\altp}{\mathsf{Assign\_Transaction\_Price}}

 \newcommand{\ts}{\mathsf{timestamp}}
 \newcommand{\pr}{\mathsf{price}}
 \newcommand{\tpr}{\mathsf{tprice}}
 \newcommand{\Q}{\mathsf{Qty}}
 \newcommand{\q}{\mathsf{qty}}
 \newcommand{\id}{\mathsf{id}}

 \newcommand{\idb}{\mathsf{id_{bid}}}
 \newcommand{\ida}{\mathsf{id_{ask}}}
 \newcommand{\idsb}{\mathsf{ids_{bid}}}
 \newcommand{\idsa}{\mathsf{ids_{ask}}}
 \newcommand{\ids}{\mathsf{ids}}
 \newcommand{\vol}{\mathsf{Vol}}

\usepackage[colorinlistoftodos,prependcaption,textsize=scriptsize]{todonotes}

\begin{document}
\title{Double Auctions: Formalization and Automated Checkers}
\author{Mohit Garg\thanks{Indian Institute of Science, Bengaluru, India. mohitgarg@iisc.ac.in. Supported by a Walmart fellowship.} 
\ \quad
N. Raja\thanks{Tata Institute of Fundamental Research, Mumbai, India. raja@tifr.res.in.}
\ \quad
 Suneel Sarswat\thanks{Tata Institute of Fundamental Research, Mumbai, India. suneel.sarswat@gmail.com.}
 \ \quad
 Abhishek Kr Singh\thanks{National University of Singapore, Singapore. abhishek.uor@gmail.com.}}

\date{}

\maketitle

\abstract{Double auctions are widely used in financial markets, such as those for stocks, derivatives, currencies, and commodities, to match demand and supply. Once all buyers and sellers have placed their trade requests, the exchange determines how these requests are to be matched. The two most common objectives for determining the matching are maximizing trade volume at a uniform price and maximizing trade volume through dynamic pricing. Prior research has primarily focused on single-quantity trade requests. In this work, we extend the framework to handle multiple-quantity trade requests and present fully formalized matching algorithms for double auctions, along with their correctness proofs. We establish new uniqueness theorems, enabling automatic detection of violations in exchange systems by comparing their output to that of a verified program. All proofs are formalized in the Coq Proof 
Assistant, and we extract verified OCaml and Haskell programs that could serve as a resource for exchanges and market regulators. We demonstrate the practical applicability of our work by running the verified program on real market data from an exchange to automatically check for violations in the exchange algorithm.}

\maketitle

\section{Introduction}
Computer algorithms are routinely deployed nowadays by all big stock exchanges
to match buy and sell requests. These algorithms are required to comply with
various regulatory guidelines. 
For example, it is mandatory for a matching resulting from
\textit{call auctions}, a type of double auction, to be \textit{fair, uniform,} and \textit{individual-rational}.

In this paper, we introduce a formal framework for analyzing trades
resulting from double auctions used in financial
markets. To verify the essential properties required by market regulators, 
we formally define these notions in the functional setting of a theorem prover and then
develop important results about matching demand and supply. Finally, we
use this framework to verify properties of two important classes of double auctions.

One of the resulting advantages of our work for an exchange or a regulator is that they can automatically check the currently deployed matching programs for any violations of required properties.
All the definitions and results presented in this
paper are completely formalized in the Coq Proof Assistant.
The complete formalization in Coq facilitates automatic program
extraction in OCaml and Haskell, with the guarantee that extracted programs satisfy
the requirements specified by the market regulator. 
Consequently, the extracted program could be deployed directly at an exchange, in addition to being used for checking violations in existing programs. We demonstrate the practical applicability of our work by running the verified program on real market data from an exchange to automatically check for violations in the exchange algorithm.

To describe our contributions in detail, we first need to provide an overview of double auctions.

\subsection{Overview of Double Auctions}
\label{sec:overview}
Financial trades occur at various types of exchanges, such as those for stocks, derivatives, commodities, and currencies. At any exchange, multiple buyers and sellers participate in trading specific products. Most exchanges use double auctions to match buyers and sellers. Additionally, some exchanges use an online continuous algorithm to execute trades during certain time intervals, particularly for highly traded products.

To conduct trades of a certain product using a double auction, the exchange collects buy and sell requests from traders over a fixed time period. At the end of this period, the exchange matches some of these requests and generates a \textit{matching}, which is a set of \textit{transactions}.
A buyer submits a buy request, also called a \emph{bid}, specifying the maximum quantity of units they wish to buy and a maximum price per unit that they can pay. Similarly, a seller submits a sell request, known as an \emph{ask}, specifying a quantity and a minimum price per unit. Each \emph{transaction} consists of a bid, an ask, a transaction quantity, and a transaction price. Naturally, the transaction quantity should be at most the minimum of the bid and ask quantities, and the transaction price should fall between the bid and ask prices, which is referred to as the individual-rational property.

Apart from the individual-rational property, 
there are other desired properties for the matching to possess. These properties are: \emph{uniform}, \emph{fair}, \emph{maximum}, and \emph{optimal uniform}. We briefly describe these matching properties:
\begin{itemize}
\item \textbf{Uniform}: A matching is uniform if all its transaction prices are the same.
\item \textbf{Fair}: A bid $b_1$ is more competitive than a bid $b_2$ if $b_1$ has a higher maximum price than $b_2$ or if their maximum prices are the same and $b_1$ arrives earlier than $b_2$. Similarly, we can define competitiveness between two asks. A matching is \emph{unfair} if a less competitive bid gets matched but a more competitive bid is not fully matched. Similarly, it could be unfair if a more competitive ask is not fully matched. If a matching is not unfair, then it is fair. 
\item \textbf{Maximum}: A matching is maximum if it has the largest volume (sum of the transaction quantities of all the transactions in the matching) among all possible matchings.
\item \textbf{Optimal-uniform}: A uniform matching is called optimal-uniform if it has the largest volume among all matchings that are uniform.
\end{itemize}
It is known that there are sets of bids and asks for which there is no possible matching that possesses all the above properties simultaneously (see, for example, \cite{WWW98,mcafee1992}).
This gives rise to two classes of double auctions, each with a different objective:
\begin{enumerate}
\item \textbf{Optimal-uniform matching}: In the context of financial markets, call auctions are commonly utilized where the matching needs to be fair and optimal-uniform, thus compromising on the maximum property. The common price in a matching produced by call auctions is often referred to as the equilibrium price, and the process is known as price discovery. 
\item \textbf{Maximum matching}: In other contexts where the matching being maximum is important, the matching needs to be fair and maximum, compromising on the uniformity property. Such double auctions are said to have dynamic-pricing.
\end{enumerate}
In our work, we consider both these classes of double auctions.

\subsection{Our Contributions}
\label{sec:contribution}
We now describe the results obtained in this work. 
For each result, apart from describing the result informally, we also provide the formal statement which involves terms that will be formally defined later. 

\begin{itemize}
\item{\textbf{Combinatorial result.}} We show that the modeling and the libraries we created to obtain our results are also useful in proving other important results on double auctions.
For example, in Theorem \ref{thm:boundM}, we prove a well-known result (Lemma $4$ of \cite{NiuP13}): for any price $p$, no matching can achieve a volume higher than the sum of the total demand and the total supply in the market at price $p$. Here the total demand at a price $p$ is the sum of the quantities of the bids whose transaction price is at least $p$. Similarly, we can define total supply. Formally,
\begin{restatable}[Demand-Supply Inequality]
{theorem}{boundM}\label{thm:boundM}
 If $M$ is a matching over an admissible order-domain $(B, A)$, then for all natural numbers $p$, we have
 $\vol(M) \leq \vol(B_{\geq p}) + \vol(A_{\leq p})$.
\end{restatable}

Here an admissible order-domain $(B, A)$ refers to a set of bids $B$ and a set of asks $A$ with unique ids and unique timestamps. In our Coq formalization, this theorem appears as follows.

\begin{minted}[bgcolor=gray!05, fontsize=\scriptsize]{coq} 
Theorem Bound p M B A: 
        admissible B A /\ Matching M B A -> 
        
        Vol(M) <= (Qty_orders (filter (fun x => p <= (oprice x)) B)) +
                  (Qty_orders (filter (fun x => p => (oprice x)) A)).
(* Here Qty_orders(L) is the sum of the quantities of the orders in L. *)
\end{minted}

\item{\textbf{Fairness.}} We show that any matching can be converted into a fair matching without compromising on the total volume. For this, we design an algorithm, the $\Fair$ procedure, which takes a matching $M$ as input, and outputs a matching $M'$.
In Theorem \ref{thm:Faircorrect}, we show that the volume of $M$ and $M'$ are the same and $M'$ is a fair matching. Formally,

\begin{restatable}[Correctness of $\Fair$]{theorem}{Faircorrect}\label{thm:Faircorrect}
If $M$ is a matching over an admissible order-domain $(B, A)$, then the matching $M' = \Fair (M ,B ,A)$ over $(B,A)$  is a fair matching such that $\vol(M) = \vol(M')$.
\end{restatable}

This theorem appears in our Coq formalization as follows.

\begin{minted}[bgcolor=gray!05, fontsize=\scriptsize]{coq} 
(* Correctness of Fair proccedure *)
Theorem Fair_main (M: list transaction) (B A: list order): 
        admissible B A /\ Matching M B A -> 
        
        (Matching (Fair M B A) B A) /\  
        (* (Fair M B A) is a matching over (B, A) *)
        (Vol(M)= Vol((Fair M B A))) /\  
        (* Trade volumes of M and (Fair M B A) are the same *)
        (Is_fair (Fair M B A) B A).     
        (* Process Fair produces a fair matching *)
\end{minted}

\item{\textbf{Uniform matching.}} We design an algorithm, the $\UM$ procedure, that takes as input a collection of bids and asks and outputs a fair and uniform matching. Furthermore, in Theorem \ref{thm:UMcorrect}, we show that the output matching
has the largest total trade volume among all the matchings that are uniform and thus is optimal-uniform. 
This algorithm implements the call auction that is used at various exchanges. Formally,

\begin{restatable}[Correctness of $\UM$]{theorem}{UMcorrect}\label{thm:UMcorrect} 
Given an admissible order-domain $(B, A)$, $\UM(B ,A)$ outputs a fair and optimal-uniform matching over $(B,A)$.
\end{restatable}

This theorem appears in our Coq formalization as follows.

\begin{minted}[bgcolor=gray!05, fontsize=\scriptsize]{coq} 
(* The UM is fair and optimal uniform algorithm. *)
Theorem UM_correct B A:
        admissible B A -> 
        Is_fair (UM B A) B A /\ Is_optimal_uniform (UM B A) B A.
\end{minted}

\item{\textbf{Maximum matching.}}  We design an algorithm, the $\MM$ procedure, that takes as input a collection of bids and asks and outputs a fair and maximum matching. In Theorem \ref{thm:MMcorrect}, we show that the output matching 
has the largest volume among all the matchings. Formally,

\begin{restatable}[Correctness of $\MM$]{theorem}{MMcorrect}\label{thm:MMcorrect}
Given an admissible order-domain $(B, A)$, $\MM(B,A)$ outputs a maximum volume matching over $(B, A)$ that is also fair.
\end{restatable}

This theorem appears in our Coq formalization as follows.

\begin{minted}[bgcolor=gray!05, fontsize=\scriptsize]{coq} 
(* The MM is fair and maximum volume matching algorithm. *)
Theorem MM_correct B A:
        admissible B A -> 
        Is_maximum (MM B A) B A /\ Is_fair (MM B A) B A.
\end{minted}

\item{\textbf{Uniqueness theorems.}}  For any two fair and optimal-uniform matchings, Theorem~\ref{thm:uniquenessTheorem} implies that for each order its total traded quantity in the two matchings is the same. Thus, if we compare the trade volumes between an exchange's matching output with our verified program's output and for some orders they do not match, then the exchange's matching is not fair and optimal-uniform. Conversely, if for each order, the trade volumes match, then Theorem~\ref{thm:ConverseUniquenessTheorem} implies that the exchange's matching is also fair and optimal-uniform (given that it already a uniform matching, which can be easily verified by checking the transaction prices). 
Making use of these results, in~\cref{sec:demonstration}, we check violations automatically in real data
from an exchange.

Formally,

\begin{restatable}[Completeness]{theorem}{uniquenessTheorem} \label{thm:uniquenessTheorem}
Let $M_1$ and $M_2$ be two fair matchings over an admissible order domain $(B, A)$ such that $\vol(M_1) = \vol(M_2)$, then for each order $\omega$, the total traded quantity of $\omega$ in $M_1$ is equal to the total traded quantity of $\omega$ in $M_2$.
\end{restatable}

\begin{restatable}[Fairness Certificate]{theorem}{ConverseUniquenessTheorem} \label{thm:ConverseUniquenessTheorem}
Given a list of bids $B$ and a list of asks $A$, if $M_1$ is a fair matching and $M_2$ is an arbitrary matching such that for each order $\omega$, the total traded quantity of $\omega$ in $M_1$ is equal to the total traded quantity of $\omega$ in $M_2$, then $M_2$ is fair.
\end{restatable}

These theorems appear in our Coq formalization as follows.
\begin{minted}[bgcolor=gray!05, fontsize=\scriptsize]{coq} 
(* Uniqueness preperty (completeness) *)
Theorem completeness M1 M2 B A:
        admissible B A /\ (Vol(M1) = Vol(M2)) /\
        (Matching M1 B A) /\ (Matching M2 B A) /\
        Is_fair M1 B A /\ Is_fair M2 B A ->

        (forall a, Qty_ask M1 (id a) = Qty_ask M2 (id a)) /\
        (forall b, Qty_bid M1 (id b) = Qty_bid M2 (id b)).

(* Converse uniqueness preperty *)
Theorem soundness M1 M2 B A:
        admissible B A /\ 
        (Matching M1 B A) /\ (Matching M2 B A) /\
        Is_fair M2 B A /\ (Vol(M1) = Vol(M2)) /\
        (forall a, Qty_ask M1 (id a) = Qty_ask M2 (id a)) /\
        (forall b, Qty_bid M1 (id b) = Qty_bid M2 (id b)) -> 

        Is_fair M1 B A.
\end{minted}

The above two theorems do not just help in building automated checkers for exchanges that output optimal-uniform matchings, but can similarly be utilized to build automated checkers for exchanges that output maximum matchings.
\end{itemize}

The Coq code together with the extracted OCaml and Haskell programs for all the
above results is available at~\cite{git:call}. Our Coq formalization consists of approximately 450 lemmas and theorems and 9000 lines of code.

\subsection{Related Work}
\label{sec:related}
In their influential work~\cite{PI17}, Passmore and Ignatovich emphasized the importance of formal verification for financial algorithms and identified several open problems in the field. In response, they developed Imandra~\cite{imandra}, a specialized formal verification system and programming language designed to reason about algorithmic properties that can be proved, refuted, or described.

Wurman, Walsh, and Wellman discuss the theory and implementation of call auctions in \cite{WWW98}. The fairness, uniform, maximum, and optimal-uniform properties described above are discussed in the works of Zhao, Zhang, Khan, and Perrussel~\cite{zhao2010maximal} and Niu and Parsons~\cite{NiuP13}. The proofs and mechanisms discussed in these papers are of single unit and not formalized. Besides this, many proofs presented in \cite{NiuP13,zhao2010maximal} are existential in nature. 

In an earlier work \cite{SS20}, Sarswat and Singh dealt primarily with single unit trade requests
and thus provided a proof of concept for obtaining verified programs for financial markets.
In this work, we generalize their results to multiple units that results in verified
programs which can be directly used in real markets. Our proofs are constructive and based on induction, which makes it easier to formalize. Furthermore, the uniqueness theorems and the resulting automated checkers that we present is a completely new contribution of this work. 

As mentioned earlier, certain exchanges also use an online continuous algorithm to match buy and sell requests. The theory, formalization, and complexity of such auctions have been studied by Garg and Sarswat in~\cite{GS22,kalpa}. Cervesato, Khan, Reis, and \v{Z}uni\'c \cite{clf} use concurrent linear logic (CLF) to outline two important properties of a 
continuous trading system.

There are also some works formalizing 
various concepts from auction theory \cite{KP18,nash,frank}, particularly 
focusing on the Vickrey auction mechanism. 

In this work, we have significantly enhanced the formalization compared to its preliminary version presented in~\cite{RSS21}. The definitions and algorithms have been streamlined, and the running time of our formalized algorithms has improved from $O(n^2)$ to $O(n \log n)$ (where $n$ denotes the number of trade requests), which enhances the practical applicability of our work.

\subsection*{Organization of the Paper}
The rest of this paper is organized as follows: In \cref{sec:modeling}, we begin with the definitions
of the various terms related to double auctions.
We then prove the demand-supply inequality (\cref{thm:boundM}) in 
 \cref{sec:demand-supply}. In \cref{sec:fair}, we describe the $\Fair$ procedure and establish its correctness (\cref{thm:Faircorrect}).
 Next, in \cref{sec:match}, we describe the $\UM$ and $\MM$ procedures and establish their correctness (\cref{thm:UMcorrect,thm:MMcorrect}). 
  Finally, in \cref{sec:uniqueness}, we prove the uniqueness theorems (\cref{thm:uniquenessTheorem,thm:ConverseUniquenessTheorem})  and explain how they give rise to automated checkers for double auctions. A practical demonstration of such a checker is also included in \cref{sec:demonstration}.

We have written our proofs and algorithms in a style that should be accessible to a broader mathematical audience, including those unfamiliar with concepts of formalization or functional programming. The functional implementation of these algorithms is available in our formalization~\cite{git:call}.

\section{Preliminaries}
\label{sec:modeling}

 In this section, we introduce the various definitions underlying our formalization of double auctions.
Many definitions in this work are analogous to the definitions introduced for continuous double auctions in~\cite{GS22}.
Our presentation leverages set notation for clarity, with the understanding that all sets discussed are finite. It is worth noting that our Coq formalization employs lists rather than sets. For brevity and intuition, we apply set-theoretic notation (e.g., $\in$, $\subseteq$, $\supseteq$, $\emptyset$) to lists, with their meanings easily inferable from context. The decision to use lists in our formalization serves two purposes: it aligns naturally with our auction modeling, and crucially, it facilitates algorithm optimization, yielding efficient implementations.

\subsection{Orders}

Unlike a previous work~\cite{RSS21}, here we adopt a unified approach to bids and asks by representing both as {\bf orders}, eliminating redundant proofs of shared properties. An order $\omega$ is defined as a 4-tuple $(id, \ timestamp, \ quantity, \ price)$, where each component—denoted as $\id(\omega)$, $\ts(\omega)$, $\q(\omega)$, and $\pr(\omega)$ respectively—is a natural number. Additionally, we stipulate that $\q(\omega) > 0$. It is important to note that prices are expressed as natural numbers, corresponding to the smallest monetary unit (e.g., cents in the United States).

\begin{minted}[bgcolor=gray!07, fontsize=\scriptsize]{coq} 
(* Definition of Order in Coq. The term 'nat' stands for natural number*)
Record order := Make_order
            {id: nat; otime: nat; oquantity: nat; oprice: nat; 
            oquantity_cond: Nat.ltb oquantity 1 = false }.
(* The term (Nat.ltb x y) checks if x < y. *)
\end{minted}

For a set of orders $\Omega$, we define $\ids(\Omega)$ as the collection of all order ids that are in $\Omega$. For a set of orders $\Omega$ with unique ids and an order $\omega \in \Omega$ where $\id(\omega) = id$, we introduce the following syntactic sugar: $\ts(\Omega,id)=\ts(\omega)$, $\q(\Omega,id)=\q(\omega)$, and $\pr(\Omega, id)=\pr(\omega)$. 
This shorthand is a slight abuse of notation, but enhances readability in subsequent discussions.

We now introduce the notion of an order-domain, the universe of bids and asks in a given context, for example, the list of bids and asks that are provided as input to a matching algorithm. We call $(B,A)$ to be an order domain if $B$ and $A$ are sets of orders.  Here, the first component, $B$, represents the set of bids, while the second component, $A$, represents the set of asks. We further define an {\bf admissible} order-domain as one where each order has a unique id and timestamp. In the context of double auctions, any input to our algorithms consisting of sets of bids and asks  will invariably form an admissible order-domain.

\begin{minted}[bgcolor=gray!07, fontsize=\scriptsize]{coq} 
Definition admissible B A := 
            (NoDup (ids B))/\(NoDup (ids A))/\
            (NoDup (timesof B))/\(NoDup (timesof A)). 
(*Note: NoDup is predicate for duplicate-free and 
(timesof B) gives timestamp's of B*)
\end{minted}

Let us now formalize the concepts of 'tradable' and 'matchable'. Consider two orders: a bid $b$ and an ask $a$. We define these orders as {\bf tradable} if the bid price meets or exceeds the ask price, i.e., $\pr(b) \geq \pr(a)$. Extending this notion, we characterize an order-domain as {\bf matchable} if it contains a bid and an ask that are tradable.

\begin{minted}[bgcolor=gray!07, fontsize=\scriptsize]{coq} 
Definition tradable b a := (oprice b >= oprice a).

Definition matchable (B A : list order):= 
            exists b a, (In a A)/\(In b B)/\(tradable b a). 
            (* Term 'In' indicates membership *)
\end{minted}

We now introduce the concept of competitiveness among orders. For bids, we define a bid $b_1$ as more {\bf competitive} than another bid $b_2$, denoted as $b_1 \succ b_2$, under two conditions:
\begin{enumerate}
    \item The price of $b_1$ exceeds that of $b_2$ ($\pr(b_1) > \pr(b_2)$), or
    \item The prices are equal, but $b_1$ has an earlier timestamp ($\pr(b_1) = \pr(b_2)$ and $\ts(b_1)$ $ < \ts(b_2)$).
\end{enumerate}
Analogously for asks, we define an ask $a_1$ as more {\bf competitive} than $a_2$, denoted as $a_1 \succ a_2$, if:
\begin{enumerate}
    \item The price of $a_1$ is lower than that of $a_2$ ($\pr(a_1) < \pr(a_2)$), or
    \item The prices are equal, but $a_1$ has an earlier timestamp ($\pr(a_1) = \pr(a_2)$ and $\ts(a_1)$ $ < \ts(a_2)$).
\end{enumerate}
The notion of competitiveness will be useful while formalizing the priority rule for double auctions.

\begin{minted}[bgcolor=gray!07, fontsize=\scriptsize]{coq} 
Definition bcompetitive b b' := 
            ((oprice b') < (oprice b)) ||                                     
            (((oprice b') == (oprice b)) && ((otime b) <= (otime b'))).       
                                                                  
Definition acompetitive a a' := 
            ((oprice a) < (oprice a')) ||                                          
            (((oprice a) == (oprice a')) && ((otime a) <= (otime a'))).
\end{minted}

\subsection{Transactions and Matchings}

A transaction is a 4-tuple $(id_b, \ id_a, \ quantity, \ price)$, where all components are natural numbers,  $id_b$ and $id_a$ represents the ids of the participating bid and ask, respectively, the $quantity$ specifies the transaction quantity, and the $price$ denotes the transaction price.
We impose the constraint that $quantity > 0$ to ensure that the transactions are meaningful. For a transaction $t$, we represent its four components by $\idb(t)$, $\ida(t)$, $\q(t)$, and $\tpr(t)$, respectively.

\begin{minted}[bgcolor=gray!07, fontsize=\scriptsize]{coq} 
Record transaction :=  Make_transaction 
        {idb: nat; ida: nat; tquantity: nat; tprice: nat; 
        tquantity_cond: Nat.ltb tquantity 1 = false }. 
\end{minted}

\begin{remark*}
Both order and transaction are record types in our Coq formalization. In our Coq definitions of these terms, oquantity\_cond and tquantity\_cond ensure that no orders or transactions are allowed with a quantity of zero. This restriction helps keeping the result statements concise; for otherwise, we would have to add a condition stating these quantities are positive in our results.
\end{remark*}

Let $T$ denote a set of transactions.
 We define $\idsb(T)$ and $\idsa(T)$ as the set of participating bid ids and ask ids in $T$, respectively. Furthermore, we define three quantities.
\begin{enumerate}
    \item $\Q_{\text{bid}}(T, id_b)$: The sum of the transaction quantities of transactions in $T$ where the participating bid has id $id_b$.
    \item $\Q_{\text{ask}}(T, id_a)$: The sum of the transaction quantities of transactions in $T$ where the participating ask has id $id_a$.
    \item $\Q_{\text{transaction}}(T, id_b \leftrightarrow id_a)$: The sum of the transaction quantities of transactions in $T$ where the participating bid and ask have ids $id_b$ and $id_a$, respectively.
\end{enumerate}
For ease of readability, we simply use $\Q$ to represent the above quantities, where the exact meaning can be easily inferred from context. We now define $\vol(T)$ as the sum of the transaction quantities of all transactions in $T$, and extend this notation to a set or orders $\Omega$: $\vol(\Omega)$ represents the sum of the maximum quantities of the orders in $\Omega$.

\begin{minted}[bgcolor=gray!07, fontsize=\scriptsize]{coq} 
(* Functional and propositional definitions of ids_bid *)
Definition fun_ids_bid T := uniq (map idb T).

(* Below, I represents ids of bids participating in T *)
Definition ids_bid I T := 
            (forall i, In i I ->(exists t, (In t T)/\(idb t = i))) /\
            (forall t, In t T ->(exists i, (In i I)/\(idb t = i))) /\
            (NoDup I).                                                  
            (* Similarly, we define ids_ask *)
\end{minted}

\begin{minted}[bgcolor=gray!07, fontsize=\scriptsize]{coq} 
(*Definition of Qty_bid and Qty_ask*)
Definition Qty_bid T i := 
           sum (map tquantity (filter (fun t => (idb t) == i) T)).
(* Sum of transaction quantities for all those transactions 
   whose bid id is equal to i. *)
(* Similarly, we define Qty_ask *)

Definition Qty_transaction T j i := 
           sum (map tquantity (filter (fun t => ((idb t) == j) &&
           (ida t) == i)) T)
\end{minted}

\begin{minted}[bgcolor=gray!07, fontsize=\scriptsize]{coq} 
(*Definition of Vol*)
Definition Vol T := sum (map tquantity T).
\end{minted}

A transaction $t$ is said to be {\bf over} an order-domain $(B,A)$ if its participating bid and ask come from $(B,A)$, i.e., $\idb(t)$ $=$ $\id(b)$ for some bid $b\in B$ and $\ida(t)=\id(a)$ for some ask $a\in A$.

A transaction $t$ is said to be {\bf valid} with respect to an order-domain $(B,A)$ if there exists bid $b\in B$ and ask $a\in A$ satisfying: 

\begin{enumerate}
    \item $\idb(t) = \id(b)$ and $\ida(t) = \id(a)$
    \item $b$ and $a$ are tradable
    \item $\q(t) \leq \min(\q(b), \q(a))$
    \item $\pr(a) \leq \tpr(t) \leq \pr(b)$
\end{enumerate}
Note that condition 4 implies condition 2, but we keep both for clarity.
We say that a set of transactions $T$ is {\bf valid} over $(B,A)$ if each of its transactions is valid over $(B,A)$.

\begin{minted}[bgcolor=gray!07, fontsize=\scriptsize]{coq} 
Definition Tvalid T B A := 
           forall t, (In t T) -> (exists b a, (In a A)/\(In b B)/\
           (idb t = id b)/\(ida t = id a)/\
           (tradable b a)/\
           (tquantity t <= oquantity b)/\(tquantity t <= oquantity a)/\
           (oprice b >= tprice t)/\(tprice t >= oprice a)).
\end{minted}

We are now ready to define a matching, that represents a feasible set of transactions that can arise from a given order-domain. We define a {\bf matching} $M$ over an admissible order-domain $(B,A)$ as a set of valid transactions where for each order $\omega \in B \cup A$, $\Q(M, \id(\omega)) \leq \q(\omega)$.

\begin{minted}[bgcolor=gray!07, fontsize=\scriptsize]{coq} 
Definition Matching M B A := 
           (Tvalid M B A)/\
           (forall b, In b B -> (Qty_bid M (id b)) <= (oquantity b))/\
           (forall a, In a A -> (Qty_ask M (id a)) <= (oquantity a)).
\end{minted}

\subsection{Classes of Matchings}
\label{sec:TypesOfMatchings}
Now we defines specific matchings relevant to call auctions.

A matching $M$ over $(B, A)$ is called a {\bf fair matching} if for each order $\omega$ that gets traded in $M$, all orders that are more competitive than $\omega$ are fully traded in $M$. Formally,
    \begin{align*}
    \text{a. } \ \forall &b,b'\in B, \ b \succ b' \text{ and } \id(b') \in \idsb(M) 
    \implies \Q(M,id(b)) = \q(b). \\
    \text{b. } \ \forall &a,a'\in A, \ a \succ a' \text{ and } \id(a') \in \idsa(M) 
    \implies \Q(M,id(a)) = \q(a) 
    \end{align*}
In the above definition, the first property is known as fair on bids and the second property is known as fair on asks. A fair matching is fair on the bids as well as fair on the asks.
\begin{minted}[bgcolor=gray!05, fontsize=\scriptsize]{coq} 
Definition Is_fair_bids M B :=
           forall b b', (In b B) /\ (In b' B) /\
           (bcompetitive b b' /\ ~eqcompetitive b b') /\   
           (* b is more competitive than b' *)
           (In (id b') (ids_bid_aux M)) ->                 
           (* b' participates in M *)
           (Qty_bid M (id b)) = (oquantity b).             
           (* b is fully traded in M *)

Definition Is_fair_asks M A :=
           forall a a', (In a A) /\ (In a' A) /\ 
           (acompetitive a a' /\ ~eqcompetitive a a') /\
           (In (id a') (ids_ask_aux M)) ->
           (Qty_ask M (id a)) = (oquantity a).

Definition Is_fair M B A := 
           Is_fair_bids M B /\ Is_fair_asks M A.  
           (* M is fair over (B, A). *)
\end{minted}

Note that, as we will see later in \cref{sec:fair}, for a given matching $M$ over $(B, A)$ there always exists a matching $M'$ over $(B, A)$ such that $M'$ is fair and $\vol(M)=\vol(M')$. 


A matching $M$ over $(B, A)$ is called a {\bf maximum matching} if it has the highest volume among all matchings over $(B,A)$, i.e., for all matchings $M'$ over $(B,A)$, $\vol(M) \ge \vol(M')$.

\begin{minted}[bgcolor=gray!05, fontsize=\scriptsize]{coq} 
Definition Is_max M B A := Matching M B A ->
           forall M', Matching M' B A /\ Vol(M) >= Vol(M').
\end{minted}

Note that there can be multiple maximum matchings over an order-domain. In Section~\ref{sec:maximum} we will see an algorithm that takes an order-domain as input and outputs a maximum matching over it, which is also fair.

Assigning different transaction prices for the same product at the same point in time might make some traders unhappy. Consequently, it is desirable that all the transactions have the same transaction price. A matching where each transaction price is the same is called a {\bf uniform matching}.

\begin{minted}[bgcolor=gray!05, fontsize=\scriptsize]{coq} 
Definition Is_uniform M B A := (Uniform M /\ Matching M B A). 
(* Here Uniform is an inductive predicate that checks 
if the trade prices of M are all equal. *)
\end{minted}

\begin{figure}[ht]
\begin{center}

\resizebox {.5\textwidth} {!} {

\begin{tikzpicture}[font=\Large]

 \draw [fill=gray!5] rectangle (22/14,2);

 \draw (.5*22/14, .5) node {$85$};

 \draw (.5*22/14, 1.5) node {$100$};

 \draw [fill=gray!5] (2*22/14,0) --(2*22/14,2)-- (3*22/14,2)-- (3*22/14,0)--cycle;

 \draw (2.5*22/14, .5) node {$90$};

 \draw (2.5*22/14, 1.5) node {$70$};

 \draw [fill=gray!5] (6*22/14,0) --(6*22/14,2)-- (7*22/14,2)-- (7*22/14,0)--cycle;

 \draw (6.5*22/14, .5) node {$85$};

 \draw (6.5*22/14, 1.5) node {$100$};

 \draw [fill=gray!5] (8*22/14,0) --(8*22/14,2)-- (9*22/14,2)-- (9*22/14,0)--cycle;

 \draw (8.5*22/14, .5) node {$90$};

 \draw (8.5*22/14, 1.5) node {$70$};

 \path (.5*22/14, 2.2) node[above] {$B$};

 \path (2.5*22/14, 2.2) node[above] {$A$};
sex
 \path (6.5*22/14, 2.2) node[above] {$B$};

 \path (8.5*22/14, 2.2) node[above] {$A$};

  \draw (1.5*22/14, -1) node[above] {(a) Uniform Matching};
  \draw (7.5*22/14, -1) node[above] {(b) Maximum Matching};


  \draw [densely dotted] (0.75*22/14,1.5) -- (2.25*22/14,1.5);

  \draw [densely dotted] (6.75*22/14,1.5) -- (8.25*22/14,0.5);

  \draw [densely dotted] (6.75*22/14,0.5) -- (8.25*22/14,1.5);

\end{tikzpicture}
}
\end{center}
\caption{Sometimes to maximize the total trade volume, we have to accept different trade prices to the matched bid-ask pairs. In this example the only matching of size two is not uniform. Here the bids (B) and the asks (A) all have quantity one each, and their limit prices are displayed.}
\label{fig:mmum}

\end{figure}
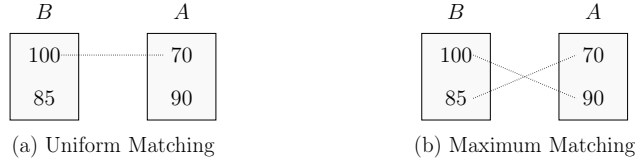

A matching $M$ over $(B, A)$ is called an {\bf optimal uniform} matching if $M$ is uniform and has the maximum volume among all the uniform matchings over $(B, A)$. 

\begin{minted}[bgcolor=gray!05, fontsize=\scriptsize]{coq} 

Definition Is_optimal_uniform M B A := Is_uniform M B A -> 
           forall M', Is_uniform M' B A /\ Vol(M) >= Vol(M').
\end{minted}

Note that the volume of an optimal uniform matching can be strictly less than the volume of a maximum matching (see \cref{fig:mmum}). Also, similar to the case of maximum matchings, there can be more than one optimal uniform matchings over a given order-domain. In Section~\ref{sec:uniform}, we exhibit an algorithm that takes as input an order-domain and outputs an optimal uniform matching, which is also fair.

\section{Demand-Supply Inequality}
\label{sec:demand-supply}
Now we are ready to present the proof of the demand-supply inequality, which provides an upper bound on the volume of an arbitrary matching in terms of the \emph{demand} and \emph{supply}. 
This is a basic inequality that is of independent interest and serves as a good warm-up before we delve into double auctions.

Given a list of bids $B$ and a list of asks $A$, where the ids are all distinct, i.e., $(B,A)$ forms an admissible domain, we first define what we mean by the total demand/supply at price $p$. To this end, let $B_{\geq p}$ represent the list of bids in $B$ whose limit prices are at least $p$ and $A_{\leq p}$ represents the list of asks in $A$ whose limit prices are at most $p$. 
The total demand at price $p$ is defined to be the sum of the quantities of orders in $B_{\geq p}$, i.e.,  $\vol (B_{\geq p})$. Similarly, the total supply at price $p$ is defined to be the sum of the quantities of the orders in $A_{\leq p}$, i.e., $\vol(A_{\leq p})$. We can now state the theorem as follows.

\boundM*

This appears in the Coq formalization as
\begin{minted}[bgcolor=gray!05, fontsize=\scriptsize]{coq} 
Theorem Bound p M B A: 
        admissible B A /\ Matching M B A -> 
        
        Vol(M) <= (Qty_orders (filter (fun x => p <= (oprice x)) B)) +
                  (Qty_orders (filter (fun x => p => (oprice x)) A)).
(* Here Qty_orders(L) is the sum of the quantities of the orders in L. *)
\end{minted}

\begin{proof}
[Proof of \cref{thm:boundM}]
First observe that the volume of any matching is upper bounded by the volume of all bids as well as the volume of all asks, i.e., if $M$ is a matching over $(B,A)$, then 

\begin{align}
\vol(M) 
\le \vol(B)\quad\text{ and } \quad\vol(M) \le \vol(A).
\end{align}

To prove the Theorem~\ref{thm:boundM}, we partition the matching $M$ into two sets: $M_1=\{(b,a,q,p') \in M \mid  \pr (b) \geq p\}$ and $M_2=\{(b,a,q,p') \in M \mid  \pr (b) < p\}$. Thus, $\vol(M) = \vol(M_1) + \vol(M_2)$. 

It is easy to see that $M_1$ is a matching over sets of bids $B_{\geq p}$ and asks $A$, and hence from the above observation, 

\begin{align}
\vol(M_1) \leq \vol(B_{\geq p}).
\end{align}

Next, we prove that $M_2$ is a matching over sets of bids $B$ and asks  $A_{< p}$. Consider a transaction $m=(b,a,q,p')$ from $M_2$. Since $m \in M$, $\pr (b) \ge \pr (a)$, and from the definition of $M_2$, we have $\pr (b) < p$. This implies $\pr(a) < p$, i.e., asks of $M_2$ come from $A_{< p}$. Hence, $M_2$ is a matching over $(B,A_{< p})$, and applying the above observation again, we have 

\begin{align}
\vol(M_2) \le \vol(A_{< p}).
\end{align}

Combining, we have 

\begin{align*}
\begin{aligned}
\vol(M) &=  \vol(M_1) + \vol(M_2) \\ 
        &\leq \vol(B_{\ge p}) + \vol(A_{< p}) \quad\quad\quad\text{ using (2) and (3)} \\ 
        &\leq \vol(B_{\ge p}) + \vol(A_{\leq p}),
\end{aligned}
\end{align*} 

which completes the proof of \cref{thm:boundM}.

\end{proof}

Formalization notes: The above proof is formalized in the file `Demand\_supply\_Inequality.v'.

\section{The $\Fair$ Algorithm}
\label{sec:fair}
In this section, we show that there exists an algorithm, which we refer to as the $\Fair$ procedure, that takes an admissible order-domain $(B, A)$ and a matching $M$ over $(B,A)$ and outputs a fair matching $M' = \Fair(M, B, A)$ over $(B,A)$ with the same volume as that of $M$, i.e., $\vol(M) = \vol(M')$. 

The $\Fair$ procedure, given an input consisting of an admissible order-domain $(B,A)$ and a matching $M$ over $(B,A)$ works in two steps: In the first step, a procedure called $\FOB$ is applied that takes  $M$ and $B$ as input and outputs a matching $M'$ over $(B,A)$ which is fair on the bids and has the same volume as that of $M$. In the second step, a procedure called $\FOA$ is applied that takes the resulting matching $M'$ and $A$ as input and outputs a matching $M''$ over $(B,A)$ which is fair on the asks and has the same volume as that of $M'$. The $\Fair$ procedure, which is the composition of the procedures $\FOA$ and $\FOB$, returns $M''$ as its output which has the same volume as $M$ and is fair (on both the bids and the asks).

The procedures $\FOB$ and $\FOA$, along with their correctness proofs, mirror each other and we just describe $\FOA$ below. We show that the $\FOA$ procedure outputs a fair on asks matching with the same volume as that of the input matching. Furthermore, if the input matching is fair on the bids, then the output matching of the $\FOA$ procedure continues to be fair on the bids. This will immediately imply that the procedure $\Fair(M,B,A)$ outputs a fair matching over $(B,A)$ with the same volume as that of $M$.

\subsection{Fair on Asks}

The $\FOA$ procedure given a matching $M$ over $(B,A)$ and the list of asks $A$, sorts the asks in $A$ in decreasing order of competitiveness (in particular, the most competitive ask is brought to the top) and sorts the transactions in the matching $M$ in increasing order of transaction prices. Then, its calls an auxiliary procedure $\FOAa$ that produces a fair on ask matching.

\begin{algorithm}[H]
\caption{The $\FOA$ Algorithm }\label{algo:foa}
\begin{algorithmic} 
\Function{$\FOA$}{Matching $M$, Asks $A$}
    \State Sort the transactions in $M$ in increasing order of its transaction prices
    \State Sort the asks in $A$ in decreasing order of competitiveness.
    \State \Return{$\FOAa(M, A, \emptyset$)}
\EndFunction    
\end{algorithmic} 
\end{algorithm}

Intuitively, when all the asks are of unit quantity, we want to scan the sorted list of the transactions $M$ from top to bottom replacing the ask ids therein with the ask ids of $A$ from top to bottom. Note that the transaction prices in $M$ will remain unchanged. This will produce a fair on asks matching. In the $\FOA$ procedure, we will implement this intuition. We just need to take care of multiple quantity asks and transactions. Furthermore, we want to make the procedure recursive so that we can provide a formalization friendly inductive proof of correctness. 

\begin{algorithm}[H]
\caption{The $\mathsf{FOA_{aux}}$ Algorithm }\label{algo:foaa}

\begin{algorithmic} 
 \Function{$\FOAa$}{Matching $M$, Asks $A$, $M_\alpha$}
\If {$|M|=0$ or $|A|=0$}
\State \Return{$M_\alpha$}
\EndIf
\State $m \leftarrow \mathsf{pop}(M)$
\State $a \leftarrow \mathsf{pop}(A)$
\State
\State $m' \leftarrow (\idb(m),\id(a), \min\{\q(m),\q(a)\}, \pr(m))$
\State $\mathsf{push}(M_\alpha, m')$
\State 
\If {$\q(m) < \q(a) $}
    \State $a \leftarrow (\id(a),\ts(a),\q(a) - \q(m),\pr(a))$
    \State $\mathsf{push}(A, a)$
\EndIf
\If {$\q(m) > \q(a)$}
    \State $m \leftarrow (\idb(m),\ida(m), \q(m) - \q(a), \pr(m))$
    \State $\mathsf{push}(M, m)$
\EndIf
\State \Return $\FOAa(M, A, M_\alpha)$
 \EndFunction    
 \end{algorithmic} 
\end{algorithm}

We initialize $M_\alpha$ to be an empty list, which at the end of the procedure will form the output. After the sorting steps, we first remove the top ask $a$ of $A$ and the top transaction $m$ of $M$. We then produce the transaction $m'$ of quantity $\min\{\q(m),\q(a)\}$ between the bid of $m$ and the ask $a$. We add $m'$ to the list $M_\alpha$.  
We then compare the quantities $\q(m)$ and $\q(a)$ to update the lists $M$ and $A$ appropriately and then recursively solve the problem on the updated lists $M$ and $A$. In the case $\q(m)=\q(a)$, we neither update $A$ nor $M$. In the case $\q(m) < \q(a)$, we reduce the quantity of ask $a$ by $\q(m)$ and insert this updated $a$ into the list $A$. Finally, in the case $\q(m)>\q(a)$, we reduce the transaction quantity of $m$ by $\q(a)$ and insert this updated $m$ into the list $M$.
  The procedure terminates when either $A$ or $M$ becomes empty. Note that since the $\vol(M) \leq \vol(A)$, $A$ cannot become empty before $M$ becomes empty. 

Having defined the $\FOA$ procedure, now we state the correctness theorem for $\FOA$. We need to show that the matching $M_\alpha$ produced by $\FOA$ is fair on the asks and the total volume of the input matching $M$ is the same as that of the output matching $M_\alpha$. We also need to show that the $\FOA$ procedure does not alter the fairness on the bids, i.e., if the input matching $M$ is fair on the bids, then the output matching $M_\alpha$ remains fair on the bids.

\begin{theorem}
\label{thm:FOA}
Let $M$ be a matching over an admissible order-domain $(B, A)$. If $M_\alpha = \FOA (M, A)$, then
the following hold.
\begin{enumerate}[label=(\alph*)]
\item $M_\alpha$ is a matching over $(B, A)$.
\item $\vol(M) = \vol(M_\alpha)$.
\item $M_\alpha$ is fair on the asks.
\item For each bid $b \in B$, $\Q(M,id(b)) = \Q(M_\alpha, id(b))$. 
\end{enumerate}
\end{theorem}
As a corollary of the last part above, we get that if $M$ is fair on the bids, then $M_\alpha$ is also fair on the bids.

The theorem statement appears in our Coq formalization as follows.
\begin{minted}[bgcolor=gray!05, fontsize=\scriptsize]{coq} 
(* The fair on ask correctness lemma. *)
Lemma FOA_correct B A:
      admissible B A /\ Matching M B A ->
      Matching (FOA M A) B A /\
      (* (a) *)
      Vol(M) = Vol(FOA M A) /\
      (* (b) *)
      Is_fair_asks (FOA M A) A /\
      (* (c) *)
      (forall b, In b B -> Qty_bid M id(b) = Qty_bid (FOA M A) id(b)).
      (* (d) *)
\end{minted}

\begin{proof}[Proof of \cref{thm:FOA}]

We give an outline of the proof, which has several obvious parts, focusing only on the most intricate aspect.

\noindent
\textbf{Proof of (a):} To prove that $M_\alpha$ is a matching, we need to show the following. 
\begin{itemize}
    \item[(i)] The ids of the bids and asks that participate in $M_\alpha$ come from $B$ and $A$, respectively.
    \item[(ii)] For each transaction $m$ in $M_\alpha$ between the bid $b$ and ask $a$, the transaction quantity of $m$ is at most $\q(b)$ and at most $\q(a)$.
    \item[(iii)] For each order $\omega$ in $B\cup A$, its total traded quantity in $M_\alpha$ is at most its total quantity $\q(\omega)$.
    \item[(iv)] For each transaction $m$ in $M_\alpha$ which is between a bid $b$ and an ask $a$, the transaction price of $m$ is between the limit prices of $b$ and $a$, i.e., $\pr(a) \leq \tpr(m) \leq \pr(b)$.
\end{itemize}

It is easy to verify that (i), (ii), (iii), and $\tpr(m) \leq \pr(b)$ of (iv) hold, as $\FOA$ always respects these constraints.
The difficult part is showing $\pr(a) \leq \tpr(m)$ of (iv), where we have to prove that when an ask id is replaced from a transaction $m$ by an ask id of an ask $a$ in $A$, the transaction price $\tpr(m)$ is at least $\pr(a)$. In other words, we need to show why the transaction prices of $M$ are respected by the replaced asks.

For ease of readability, we will not keep the list of output transactions in the argument of $\FOAa$, i.e., we will write $\FOAa(M,A)$ instead of $\FOAa(M,A,M_\alpha)$.
Let $M$ be a matching over $(B,A)$ which is sorted by increasing transaction prices and $A$ is sorted by decreasing competitiveness, i.e., the ask with the smallest price is on top of $A$. We need to show that for each transaction $m$ in $M_\alpha=\FOAa(M,A)$ if the ask participating in $m$ is $a$ then $\pr(a)\leq \tpr(m)$.
We will show this by induction on $|M|+|A|$ (note that in each recursive call of $\FOAa$ either the size of the first argument $|M|$ decreases or the size of the second argument $|A|$ decreases).
In fact, we will show a slightly general statement:

\begin{claim*} If $M$ is a matching over some admissible order-domain $(\hat B,\hat A)$ and $M$ and $A$ have the {\it supply property} (defined below), then for each transaction in $M_\alpha=\FOAa(M,A)$ whose participating ask is $a\in A$, $\pr(a)\leq \tpr(m)$. 
\end{claim*}

Note that $A$ need not be the same as $\hat A$ in the above statement. This strengthening is crucial for our proof to work, as when $\FOAa(M,A)$ makes a recursive call to $\FOAa(M',A')$, then $M'$ might have participating asks that are not present in $A'$.

\noindent\textbf{Supply property:} Let us now define the supply property. We say a matching $M$ (over an arbitrary admissible order-domain) and asks $A$ have the supply property if 
$$\vol(A_{\le p}) \geq \vol(M_{\leq p}),$$
for all transaction prices $p$ in $M$, where $A_{\leq p} = \{a\in A \mid \pr(a)\leq p\}$ and $M_{\leq p}=\{m \in M \mid \tpr(m) \leq p\}$. 

Observe that if $M$ is over $(B,A)$, then $M$ and $A$ trivially satisfy the supply property since all transactions in $M$ with transaction prices at most $p$ have participating asks from $A$ whose prices are at most $p$. Thus, showing the claim is enough to complete this part of the proof.

Let $m$ be the top transaction in $M$, i.e., with the smallest transaction price, and $a$ be the most competitive ask in $A$, i.e., the one with the smallest limit price.
Notice that when $\FOAa(M,A)$ is called it first outputs a 
transaction $m'$ of quantity $q=\min\{\q(m),\q(a)\}$, transaction price $p=\tpr(m)$, where the participating bid is the bid of $m$ and the participating ask is $a$. Here clearly, $\pr(a) \leq p=\tpr(m)$, as $a$ is the most competitive ask and there exists an ask in $A$ with price at most $p$ as from the supply property $\vol(A_{\leq p}) \geq \vol(M_{\leq p})\geq \q(m)> 0$. 

Now the remaining transactions output by $\FOAa$ are obtained from the recursive call made on $\FOAa(M',A')$, where $M'$ is obtained from $M$ by the reducing quantity $q$ from $m$ (if $q=\q(m)$, $m$ is deleted from $M$), and $A'$ is obtained from $A$ by reducing the quantity of $a$ by $q$ (if $q=\q(a)$, $A$ is deleted from $A$). As noted earlier, $|M'|+|A'|\leq |M|+|A|$. We will now be immediately done with induction. To apply the induction, however, we need to show that $M'$ and $A'$ satisfy the supply property. This is easy to deduce as initially $M$ and $A$ satisfied the supply property, and $q$ quantity was reduced from both $M$ and $A$ which had the lowest prices, at most the smallest transaction price $\pr(m)$, to obtain $M'$ and $A'$; this will imply that $M'$ and $A'$ continue to have the supply property: for each transaction price $p$ of $M'$, we have

$$ \vol(A_{\le p}) \geq \vol(M_{\leq p}) 
\implies  \vol(A_{\le p}) -q \geq \vol(M_{\leq p}) -q \implies  \vol(A'_{\le p})\geq \vol(M'_{\leq p}).$$

This completes the proof of (a). Proofs of (b), (c), and (d) follow straightforwardly. To see part (b), notice that $\vol(A)\geq \vol(M)$, and the algorithm will be able to replace every transaction in $M$ with asks in $A$. Part (c) also follows immediately, since the most competitive asks are given priority by the algorithm. Part (d) follows from observing that the bids and their quantities that participate in $M$ remain completely unaltered by $\FOA$.

\end{proof}

As explained earlier, similar to the $\FOA$ procedure, we have the $\FOB$ procedure, that produces a matching that is fair on the bids. Combining the $\FOB$ and $\FOA$ procedures, we have the following definition of the $\Fair$ procedure. $$\Fair(M ,B ,A) =  \FOA(\FOB(M, B), A)$$.

\begin{remark*}
{Note that we could have also chosen a different definition for $\Fair$. Namely, $$\Fair(M, B, A) = \FOB(\FOA(M, A),B).$$ 
    Let $M_1 = \FOA(\FOB(M, B), A)$ and $M_2 = \FOB(\FOA(M, A), B)$. Both $M_1$ and $M_2$ can be proven to be fair matchings with the same total volume. 
    Furthermore, later in \cref{sec:uniqueness} we will prove \cref{thm:uniquenessTheorem}, which, in particular, implies that for each order $\omega$, the total traded quantity of $\omega$ in $M_1$ is equal to the total traded quantity of $\omega$ in $M_2$.} 

    {Notice that in producing both $M_1$ and $M_2$, the initial matching and the intermediate matching are sorted based on transaction prices: once while applying $\FOA$ and once when applying $\FOB$. It is easy to see that if the sorting done during $\FOA$ and $\FOB$ are {\it consistent}, that is, if the second sort on the intermediate matching results in the list of transactions being reversed, then $M_1 = M_2$.
    In particular, if the transaction prices in the initial matching $M$ are all distinct, then the two sorts will always be consistent, and hence $M_1=M_2$.}
    
    {However, the matchings $M_1$ and $M_2$ may not be the same if the two sorting steps are not consistent. An easy way to see an example of this is to imagine a matching $M$ where all transaction prices are the same, and all the bids in $B$ and asks in $A$ are of unit quantity and get completely traded in $M$; in particular, $|M|=|A|=|B|$. Since all transaction prices are identical, permuting $M$ in any order is a valid sorting. Let us assume that the sorting outputs a uniformly random permutation of the input list. In that case each of $M_1$ and $M_2$ will be a uniformly random pairing of the bids $B$ and the asks $A$, and $M_1=M_2$ will occur with a probability of only $\frac{1}{|B|!}$. }
\end{remark*}

We conclude this section by formally summarizing our main result on the $\Fair$ procedure.

\Faircorrect*

This theorem statement appears in our Coq formalization as follows.  

\begin{minted}[bgcolor=gray!05, fontsize=\scriptsize]{coq} 
(* Correctness of Fair procedure. *)
Theorem Fair_main (M: list transaction) (B A: list order): 
        admissible B A /\ Matching M B A -> 
        
        (Matching (Fair M B A) B A) /\  
        (* (Fair M B A) is a matching over (B, A) *)
        (Vol(M)= Vol((Fair M B A))) /\  
        (* Trade volumes of M and (Fair M B A) are the same *)
        (Is_fair (Fair M B A) B A).     
        (* Process Fair produces a fair matching *)
\end{minted}

Formalization notes: The procedure $\FOB$ and $\FOA$ are implemented in Coq using the Equations plugin which is helpful to write functions involving well-founded recursion \cite{equation}. The proof of Theorem~\ref{thm:Faircorrect} is done in several parts. First, we prove all the parts of \cref{thm:FOA} in the file `Fair\_Ask.v'. We prove similar theorems for the procedure $\FOB$ in `Fair\_Bid.v' file. Later all the results are combined in the file `Fair.v' and the above theorem is proved as \emph{Fair\_exists}.


\section{Matching Algorithms}
\label{sec:match}

In this section, we formalize two almost identical $O(n \log n)$ time algorithms for the maximum matching and the optimal uniform matching problems. These algorithms start with sorting the list of bids in decreasing order of competitiveness. Next, the list of asks is sorted based on competitiveness: for maximum matching in increasing order, whereas, for optimal uniform matching, in decreasing order.
After the sorting step, both algorithms work in linear time using the $\match$ subroutine as follows (See Algorithm~\ref{algo:match}).  The bid $b$ on top of its sorted list is matched with the ask $a$ on top of its sorted list if they are tradable, i.e., $\pr(b) \ge \pr(a)$. 
In this case, a transaction between them is established with transaction quantity $q= \min(\q(b),q(a))$ and transaction price $\pr(a)$;\footnote{Observe that any value in the interval of the limit prices of the matched bid-ask pair can be assigned as the transaction price.} a quantity of $q$ is reduced from their existing quantities; finally, the $0$ quantity orders are deleted from the lists. 
If the orders $b$ and $a$ are not matchable, ask $a$ is deleted.
The above steps are then repeatedly applied until one of the lists becomes empty.\footnote{A symmetric version of $\match$ can be used as well, where if $b$ and $a$ are not tradable, $b$ is deleted.}
Finally, for uniform price matching, all transaction prices are replaced by the transaction price of the last transaction, which can be achieved in linear time (this step is not done by $\match$).

\begin{algorithm}[H]
\label{algo:match}
\caption{The $\match$ subroutine}
\begin{algorithmic} 
 \Function{$\match$}{Bids $B$, Asks $A$, Matching $M$}
 \Comment{Initially, $M=\emptyset$.}

\If {$|B|=0$ or $|A|=0$}
\State \Return{M}
\EndIf
 \State
\State $b\leftarrow \mathsf{pop}(B)$
\State $a\leftarrow \mathsf{pop}(A)$
\State

\If {$\pr(b)<\pr(a)$}
\State $\mathsf{push}(B, b)$
\State \Return $\match(B, A, M)$
\EndIf
\Comment{otherwise, b and a are matchable}

\State $q\leftarrow \min\{\q(a),\q(b)\}$
\State
\State $\mathsf{push}(M, \{(\id(b),\id(a), q, \pr(a))\})$
\State
\If {$\q(b) - q > 0$}
\State $\mathsf{push}(B,(\id(b),\ts(b),\q(b)-q,\pr(b)))$
\EndIf

\If {$\q(a) - q > 0$}
\State $\mathsf{push}(A,(\id(a),\ts(a),\q(a)-q,\pr(a)))$
\EndIf

\State
\State \Return $\match(B,A,M)$
 \EndFunction    
 \end{algorithmic} 
\end{algorithm}

We are going to prove the correctness of optimal uniform and maximum matching algorithms separately in the next two subsections. Since both these algorithms use the $\match$ subroutine, we first describe some important properties about $\match$ that will be used later.

We begin with observing three properties of the $\match$ subroutine.

\begin{proposition}
\label{lem:match_matching}
    If $(B, A)$ is an admissible order domain, then $\match(B,A,\emptyset)$ outputs a matching over $(B,A)$.
\end{proposition}

\begin{proposition}
\label{lem:match_fob}
    If $(B, A)$ is an admissible order-domain and $B$ is sorted in decreasing competitiveness of the bids, then $\match(B,A,\emptyset)$ outputs a matching that is fair on the bids.
\end{proposition}

\begin{proposition}
\label{lem:match_foa}
    If $(B, A)$ is an admissible order-domain and $A$ is sorted in decreasing competitiveness of the asks, then $\match(B,A,\emptyset)$ outputs a matching that is fair on the asks.
\end{proposition}

In our Coq formalization, these propositions appear as follows.

\begin{minted}[bgcolor=gray!05, fontsize=\scriptsize]{coq} 
(* The Match subroutine outputs a matching over (B, A). *)
Lemma Match_Matching B A:
      admissible B A -> Matching (Match B A) B A.
\end{minted}

\begin{minted}[bgcolor=gray!05, fontsize=\scriptsize]{coq} 
(* The Match subroutine outputs a fair on bids matching. *)
Lemma Match_Fair_on_Bids B A:
      admissible B A /\ Sorted bcompetitive B -> Is_fair_bids (Match B A) B.
\end{minted}

\begin{minted}[bgcolor=gray!05, fontsize=\scriptsize]{coq} 
(* The Match subroutine outputs a fair on asks matching. *)
Lemma Match_Fair_on_Asks B A:
      admissible B A /\ Sorted acompetitive A -> Is_fair_asks (Match B A) A.
\end{minted}

The proofs of these propositions are quite straightforward and we omit the tedious details here.

Next, we prove one of the main lemmas for the $\match$ subroutine that will be crucially used in establishing the optimality of our uniform matching algorithm in the next subsection.

\begin{lemma}
\label{lem:match_optimal_um}
    If $(B, A)$ is an admissible order-domain and bids of $B$ and asks of $A$ are sorted by decreasing competitiveness, then $M = \match(B, A,\emptyset)$ outputs a matching whose volume is at least the volume of an optimum uniform matching over $(B,A)$, i.e., for all uniform matching $M'$ over $(B, A)$, $\vol (M) \ge \vol (M') $.
\end{lemma}

In our Coq formalization, this lemma appears as follows.

\begin{minted}[bgcolor=gray!05, fontsize=\scriptsize]{coq} 
(* The Match is optimal uniform when B and A are sorted by competitiveness. *)
Theorem Match_optimal_um B A:
        admissible B A /\ 
        Sorted bcompetitive B /\ Sorted acompetitive A /\ 
        Is_uniform M B A -> 
        Vol(Match B A) >= Vol(M).
\end{minted}

To prove the above lemma, we will use the following lemma which states that if $M$ is a uniform matching over $(B, A)$ with total volume at least the minimum of quantities of the most competitive bid $b\in B$ and the most competitive ask $a\in A$, then there exists a uniform matching $M_{ab}$ of the same volume containing a transaction between $b$ and $a$ with transaction quantity precisely $\min\{\q(a),\q(b)\}$ (the maximum possible trade between $b$ and $a$).

\begin{lemma} \label{lem:surgery1} Let $b$ and $a$ be the most competitive bid and ask in $B$ and $A$, respectively.
If $M$ is a uniform matching over $(B, A)$ such that $\vol(M) \ge \min(\q(b),\q(a))$, then there exists a uniform matching $M_{ab}$ over $(B, A)$ such that $\vol(M) = \vol(M_{ab})$ and $M_{ab}$ contains a transaction between $b$ and $a$ with quantity $\min\{\q(b),\q(a)\}$. 
\end{lemma}

We first prove \cref{lem:match_optimal_um} assuming \cref{lem:surgery1}, which will be proved subsequently.

\begin{proof}[Proof of \cref{lem:match_optimal_um}]
Let $(B,A)$ be an admissible order-domain, where $B$ and $A$ are sorted by decreasing competitiveness. Let $M=\match(B,A,\emptyset)$, and let $M'$ be an arbitrary uniform matching over $(B,A)$. We need to show that $\vol(M)\geq\vol(M')$.
We prove this by induction on $(|B| + |A|)$. In the base case, $B = \emptyset$ or $A = \emptyset$, which implies $\vol(M')=0$, and we are trivially done.

\noindent
Induction step: $|B|\geq 1$ and $|A|\geq 1$. $\match$ first removes the top orders $b$ and $a$ from $B$ and $A$, respectively, and compares their prices. Since both $B$ and $A$ are sorted, $b$ is the most competitive bid of $B$ and $a$ is the most competitive ask of $A$. We have two cases: $\pr(b) < \pr(a)$ and $\pr(b) \ge \pr(a)$. In the first case, when $\pr(b) < \pr(a)$, since the most competitive bid $b$ of $B$ is not tradable with the most competitive ask $a$ of $A$, $B$ and $A$ are not matchable, which implies $\vol(M')=0$, and we are done.

In the second case, when $\pr(b) \ge \pr(a)$, the $\match$ subroutine generates a transaction with transaction quantity $q = \min (\q(b),\q(a))$, before making a recursive call. Thus, $\vol(M)\geq q$. It then generates the remaining set of transactions $\hat M$ by recursively calling $\match$ on the reduced order-domain $(\hat B, \hat A)$, which is obtained from $(B,A)$ by reducing a quantity of $q$ from each of $b$ and $a$, and deleting the zero quantity orders. In particular, at least one of $b$ and $a$ will be deleted from its respective list, and $|\hat B|+|\hat A| < |B|+|A|$. Also, $\vol(M)=q+\vol(\hat M)$.

Now, if $\vol(M')\leq q$ then we are again done as $\vol(M)\geq q$. In the case when 
 $\vol(M') \ge q$, we invoke Lemma~\ref{lem:surgery1} to obtain a uniform matching $M_{ab}$ over $(B,A)$ such that $\vol(M')=\vol(M_{ab})$ and $M_{ab}$ consists of a transaction between $b$ and $a$ with transaction quantity $q$. To complete the proof, it is sufficient to show

\begin{equation} \label{eq1} \vol(M) \ge \vol(M_{ab}). \end{equation}

 We now obtain the matching $\hat M_{ab}$ from $M_{ab}$ by deleting the transaction between $b$ and $a$ (with transaction quantity $q$). Notice $\hat M_{ab}$ is a matching over $(\hat B,\hat A)$ and

\begin{equation} \label{eq2} \vol(M_{ab}) = q + \vol(\hat M_{ab}). \end{equation}

Since $(|\hat B| + |\hat A|) < (|B| + |A|)$, $\hat M$ is over $(\hat B,\hat A)$, and $\hat M_{ab}$ is a uniform matching over $(\hat B, \hat A)$, from the induction hypothesis we have 

\begin{equation} \label{eq3} \vol(\hat M) \ge \vol(\hat M_{ab}). \end{equation}

Combining (\ref{eq2}) and (\ref{eq3}), we obtain (\ref{eq1}): 

$$ \vol(M) = q + \vol(\hat M)) \ge q + \vol(\hat M_{ab}) = \vol(M_{ab}). $$

\end{proof}

Having finished this proof, we now turn to the proof of \cref{lem:surgery1} that we assumed.

\begin{figure}[ht]
\begin{center}

\resizebox {.7\textwidth} {!} {

\begin{tikzpicture}[font=\Large]

 \draw [fill=gray!5] (0,0) -- (4,0) -- (4,3) -- (0,3)--cycle;
 \fill [fill=yellow!70] (0,.5) -- (1,.5) -- (1,1) -- (0,1)--cycle;
 \fill [fill=red!20] (0,.5+1.8) -- (1,.5+1.8) -- (1,1+1.8) -- (0,1+1.8)--cycle;
 \fill [fill=blue!20] (1,.5) -- (2,.5) -- (2,1) -- (1,1)--cycle;
 \fill [fill=green!20] (1,.5+1.8) -- (2,.5+1.8) -- (2,1+1.8) -- (1,1+1.8)--cycle;
 \draw (2, 2) node[below] {$M$};
 \draw (.5, .75+1.8) node {\small{$b$}};
 \draw (1.5, .75) node {\small{$a$}};
 \draw (1.5, .75+1.8) node {\small{$a_1$}};
 \draw (.5, .75) node {\small{$b_1$}};

 \fill [fill=white] (2,.5) -- (3,.5) -- (3,1) -- (2,1)--cycle;
 \draw (2.5, .75) node {\small{$q_2$}};

 \fill [fill=white] (3,.5) -- (4,.5) -- (4,1) -- (3,1)--cycle;
 \draw (3.5, .75) node {\small{$p$}};
 \draw [dotted] (3,.5) -- (3,1);

 \fill [fill=white] (2,.5+1.8) -- (3,.5+1.8) -- (3,1+1.8) -- (2,1+1.8)--cycle;
 \draw (2.5, .75+1.8) node {\small{$q_1$}};

 \fill [fill=white] (3,.5+1.8) -- (4,.5+1.8) -- (4,1+1.8) -- (3,1+1.8)--cycle;
 \draw (3.5, .75+1.8) node {\small{$p$}};
 \draw [dotted] (3,.5+1.8) -- (3,1+1.8);
 \draw (4,0) -- (4,3) --cycle;
 \draw (0,0) -- (0,3) --cycle;

 \draw (-.5, .75+1.8) node {$m_1$};
 \draw (-.5, .75) node {$m_2$};

 \draw [fill=gray!5,xshift=2.5in] (0,0) -- (4,0) -- (4,4) -- (0,4)--cycle;
 \fill [fill=yellow!70,xshift=2.5in] (0,.5) -- (1,.5) -- (1,1) -- (0,1)--cycle;
 \fill [fill=red!20,xshift=2.5in] (0,.5+1.8) -- (1,.5+1.8) -- (1,1+1.8) -- (0,1+1.8)--cycle;
 \fill [fill=blue!20,xshift=2.5in] (1,.5) -- (2,.5) -- (2,1) -- (1,1)--cycle;
 \fill [fill=green!20,xshift=2.5in] (1,.5+1.8) -- (2,.5+1.8) -- (2,1+1.8) -- (1,1+1.8)--cycle;
 \draw (2, 2) node[below,xshift=2.5in] {$M'$};
 \draw (.5, .75+1.8) node[xshift=2.5in] {\small{$b$}};
 \draw (1.5, .75) node[xshift=2.5in] {\small{$a$}};
 \draw (1.5, .75+1.8) node[xshift=2.5in] {\small{$a_1$}};
 \draw (.5, .75) node[xshift=2.5in] {\small{$b_1$}};

 \fill [fill=white,xshift=2.5in] (2,.5) -- (3,.5) -- (3,1) -- (2,1)--cycle;
 \draw (2.5, .75) node[xshift=2.5in] {\small{$q_2 - 1$}};

 \fill [fill=white,xshift=2.5in] (3,.5) -- (4,.5) -- (4,1) -- (3,1)--cycle;
 \draw (3.5, .75) node[xshift=2.5in] {\small{$p$}};
 \draw [dotted,xshift=2.5in] (3,.5) -- (3,1);

 \fill [fill=white,xshift=2.5in] (2,.5+1.8) -- (3,.5+1.8) -- (3,1+1.8) -- (2,1+1.8)--cycle;
 \draw (2.5, .75+1.8) node[xshift=2.5in] {\small{$q_1 - 1$}};

 \fill [fill=white,xshift=2.5in] (3,.5+1.8) -- (4,.5+1.8) -- (4,1+1.8) -- (3,1+1.8)--cycle;
 \draw (3.5, .75+1.8) node[xshift=2.5in] {\small{$p$}};
 \draw [dotted,xshift=2.5in] (3,.5+1.8) -- (3,1+1.8);
 \draw [xshift=2.5in] (4,0) -- (4,4) --cycle;
 \draw [xshift=2.5in] (0,0) -- (0,4) --cycle;

 \fill [fill=yellow!70,xshift=2.5in] (0,.5+2.5) -- (1,.5+2.5) -- (1,1+2.5) -- (0,1+2.5)--cycle;
 \fill [fill=red!20,xshift=2.5in] (0,.5+3) -- (1,.5+3) -- (1,1+3) -- (0,1+3)--cycle;
 \fill [fill=green!20,xshift=2.5in] (1,.5+2.5) -- (2,.5+2.5) -- (2,1+2.5) -- (1,1+2.5)--cycle;
 \fill [fill=blue!20,xshift=2.5in] (1,.5+3) -- (2,.5+3) -- (2,1+3) -- (1,1+3)--cycle;
 \fill [fill=white,xshift=2.5in] (2,.5+2.5) -- (3,.5+2.5) -- (3,1+2.5) -- (2,1+2.5)--cycle;
 \fill [fill=white,xshift=2.5in] (3,.5+2.5) -- (4,.5+2.5) -- (4,1+2.5) -- (3,1+2.5)--cycle;
 \fill [fill=white,xshift=2.5in] (2,.5+3) -- (3,.5+3) -- (3,1+3) -- (2,1+3)--cycle;
 \fill [fill=white,xshift=2.5in] (3,.5+3) -- (4,.5+3) -- (4,1+3) -- (3,1+3)--cycle;
 \draw [xshift=2.5in] (0,0) -- (4,0) -- (4,4) -- (0,4)--cycle;
 \draw [dotted,xshift=2.5in] (0,3) -- (4,3);
 \draw [dotted,xshift=2.5in] (0,3.5) -- (4,3.5);

 \draw [xshift=2.5in] (.5, 3.25) node {\small{$b_1$}};
 \draw [xshift=2.5in] (1.5, 3.25) node {\small{$a_1$}};
 \draw [xshift=2.5in] (2.5, 3.25) node {\small{$1$}};
 \draw [xshift=2.5in] (3.5, 3.25) node {\small{$p$}};

 \draw [xshift=2.5in] (.5, 3.75) node {\small{$b$}};
 \draw [xshift=2.5in] (1.5, 3.75) node {\small{$a$}};
 \draw [xshift=2.5in] (2.5, 3.75) node {\small{$1$}};
 \draw [xshift=2.5in] (3.5, 3.75) node {\small{$p$}};

 \draw [dotted,xshift=2.5in] (3,3) -- (3,4);

 \draw [xshift=2.5in] (-.5, .75+1.8) node {$m_1'$};
 \draw [xshift=2.5in] (-.5, .75) node {$m_2'$};
 \draw (5, 1.5) node {{$\Rightarrow$}};
\end{tikzpicture}
}
\end{center}
\caption{In the above figure the matching $M'$ is obtained from the matching $M$. Each bid or ask has the same trade quantity in both $M$ and $M'$. Furthermore, the trade quantity between $a$ and $b$ in $M'$ is one more than that in $M$. 
}
\label{fig:surgery}

\end{figure}
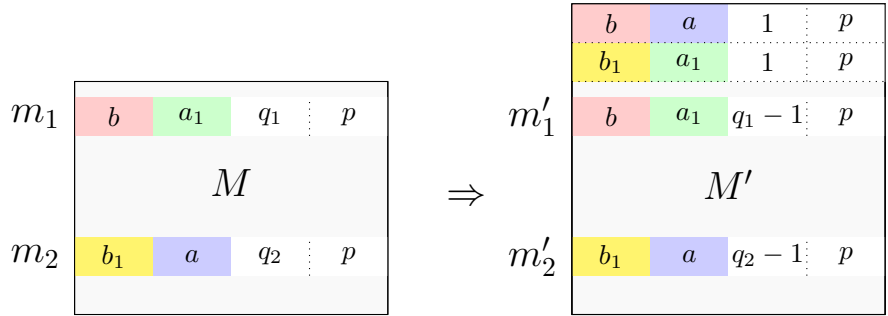

\begin{proof}
[Proof of \cref{lem:surgery1}]
Given a uniform matching $M$ with $\vol(M) \ge \min\{\q(b),\q(a)\}$ over $(B, A)$, where $b\in B$ and $a\in A$ are the most competitive bid and ask, respectively, we need to show the existence of a uniform matching $M_{ab}$ such that $\vol(M_{ab}) = \vol(M)$ and $M_{ab}$ contains a transaction between $b$ and $a$ with transaction quantity 
$\min\{\q(b),\q(a)\}$. 
Let $q = \min\{\q(b),\q(a)\}$. We do the following surgery on $M$ in two steps to obtain the desired $M_{ab}$. 

Step 1: We first modify $M$ to ensure that bid $b$ and ask $a$ each has at least $q$ total trade quantities in $M$ (not necessarily between each other). This is accomplished by running the $\Fair$ procedure on $M$ that outputs a matching that prefers the most competitive orders ($b$ and $a$) over any other orders. Since $\vol(M) \ge q$, we get that $\Fair(M,B,A)$ has at least $q$ trade quantities for each of $b$ and $a$. Note that $\Fair$ does not change the volume or affect the uniform properties of $M$. Set $M \leftarrow \Fair(M,B,A)$.

Step 2: In this step, we modify $M$ to ensure that the bid $b$ and ask $a$ have $q$ quantity traded between them. Note that in $M$ individually both $b$ and $a$ have at least $q$ total trade quantities. We will inductively transfer quantities of $b$ and $a$ that are not between them to the transaction between $b$ and $a$, a unit quantity at a time, till they have $q$ quantity trade between them. To better understand this, consider the case when $b$ and $a$ have zero trade quantity between them. Let us say there is a transaction between $b$ and $a_1$ of quantity $q_1$ and a transaction between $a$ and $b_1$ of quantity $q_2$. We remove these two transactions and replace them with the following four transactions (see Figure~\ref{fig:surgery}) that keep the matching volume intact: (1) transaction between $b$ and $a_1$ of quantity $q_1 -1$, (2) transaction between $a$ and $b_1$ of quantity $q_2 - 1$, (3) transaction between $b_1$ and $a_1$ of quantity one and (4) transaction between $b$ and $a$ of quantity one. Recall, in a uniform matching with price $p$, the limit price of each bid is at least $p$ and the limit price of each ask is at most $p$, implying any bid and ask participating in the matching are tradable. Thus, doing such a replacement surgery is legal and does not affect the uniformity property, and we obtain the desired $M'$ by repeatedly doing this surgery. 
\end{proof}

Next we prove another important lemma of $\match$ that will be useful in establishing the optimality of the maximum matching algorithm in a subsequent subsection.

\begin{lemma}
\label{lem:match_optimal_mm}
    Let $(B, A)$ be an admissible order-domain such that $B$ is sorted by decreasing competitiveness of its bids, whereas, $A$ is sorted by increasing  competitiveness of its asks. Then, $M = \match(B, A)$ outputs a maximum volume matching over $(B, A)$, i.e., for all matchings $M'$ over $(B, A)$, $\vol (M) \ge \vol (M') $.
\end{lemma}

In our Coq formalization, this lemma appears as follows.

\begin{minted}[bgcolor=gray!05, fontsize=\scriptsize]{coq} 
Theorem Match_optimal_mm B A:
        admissible B A /\ 
        Sorted bcompetitive B /\ Sorted rev_acompetitive A /\ 
        Matching M B A -> 
        Vol(Match B A) >= Vol(M).
\end{minted}

Note that this lemma is very similar to Lemma~\ref{lem:match_optimal_mm} and so is the proof. Hence, like Lemma~\ref{lem:surgery1}, we need the following lemma to prove the above result.

\begin{lemma} \label{lem:surgery2}
Let the most competitive bid of $B$ be $b$ and the least competitive ask of $A$ be $a$ such that $b$ and $a$ are tradable. If $M$ is a matching over $(B, A)$ such that $\vol(M) \ge \min(\q(b),\q(a))$, then there exists another matching $M_{ab}$ over $(B, A)$ such that $\vol(M) = \vol(M_{ab})$ and  $M_{ab}$ contains a transaction between $b$ and $a$ with quantity $\min \{\q(b),\q(a)\}$. 
\end{lemma}

We first prove \cref{lem:match_optimal_mm} assuming \cref{lem:surgery2}, which is proved subsequently.

\begin{proof}[Proof idea of Lemma~\ref{lem:match_optimal_mm}]
Let $(B, A)$ be an admissible order-domain, where $B$ is sorted by decreasing competitiveness of its bids and $A$ is sorted by increasing competitiveness of its asks. Let $M=\match(B,A,\emptyset)$, and let $M'$ be an arbitrary matching over $(B,A)$. We need to show that $\vol(M)\geq\vol(M')$. We prove this by induction on $(|B| + |A|)$. In the base case, $B = \emptyset$ or $A = \emptyset$, which implies $\vol(M')=0$, and we are trivially done.

\noindent
Induction step: $|B|\geq 1$ and $|A|\geq 1$. $\match$ first removes the top orders $b$ and $a$ from $B$ and $A$, respectively, and compares their prices. Since both $B$ and $A$ are sorted in decreasing and increasing orders of their competitiveness, respectively, $b$ is the most competitive bid of $B$ and $a$ is the least competitive ask of $A$. We have two cases: $\pr(b) < \pr(a)$ and $\pr(b) \ge \pr(a)$.

In the first case, we have $\pr(b) < \pr(a)$, i.e., $b$ and $a$ are not tradable. Since the most competitive bid $b$ is not tradable with $a$, the ask $a$ is not tradable with any bid in $B$. Therefore, both $M$ and $M'$ are matchings over $(B,A\setminus\{a\})$. As $|B|+|A\setminus\{a\}|<|B|+|A|$, we are immediately done by applying the induction hypothesis.

In the second case, when $\pr(b) \ge \pr(a)$, the $\match$ subroutine generates a transaction with transaction quantity $q = \min (\q(b),\q(a))$, before making a recursive call. Thus, $\vol(M)\geq q$. It then generates the remaining set of transactions $\hat M$ by recursively calling $\match$ on the reduced order-domain $(\hat B, \hat A)$, which is obtained from $(B,A)$ by reducing a quantity of $q$ from each of $b$ and $a$, and deleting the zero quantity orders. In particular, at least one of $b$ and $a$ will be deleted from its respective list, and $|\hat B|+|\hat A| < |B|+|A|$. Also, $\vol(M)=q+\vol(\hat M)$.

Now, if $\vol(M')\leq q$ then we are again done as $\vol(M)\geq q$. In the case when 
 $\vol(M') \ge q$, we invoke Lemma~\ref{lem:surgery2} to obtain a matching $M_{ab}$ over $(B,A)$ such that $\vol(M')=\vol(M_{ab})$ and $M_{ab}$ consists of a transaction between $b$ and $a$ with transaction quantity $q$. To complete the proof, it is sufficient to show
 
 \begin{equation} \label{eq4} \vol(M) \ge \vol(M_{ab}). \end{equation}

 We now obtain the matching $\hat M_{ab}$ from $M_{ab}$ by deleting the transaction between $b$ and $a$ (with transaction quantity $q$). Notice $\hat M_{ab}$ is a matching over $(\hat B,\hat A)$ and

\begin{equation} \label{eq5} \vol(M_{ab}) = q + \vol(\hat M_{ab}). \end{equation}

Since $(|\hat B| + |\hat A|) < (|B| + |A|)$, $\hat M$ is over $(\hat B,\hat A)$, and $\hat M_{ab}$ is a matching over $(\hat B, \hat A)$, from the induction hypothesis we have 

\begin{equation} \label{eq6} \vol(\hat M) \ge \vol(\hat M_{ab}). \end{equation}

Combining (\ref{eq5}) and (\ref{eq6}), we obtain (\ref{eq4}): 

$$ \vol(M) = q + \vol(\hat M)) \ge q + \vol(\hat M_{ab}) = \vol(M_{ab}). $$

\end{proof}

Having finished this proof, we now discuss the proof of the lemma that we assumed.
 
\begin{proof}[Proof of \cref{lem:surgery2}] 
Given a matching $M$ with $\vol(M) \ge \min\{\q(b),\q(a)\}$ over $(B, A)$, where $b\in B$ is the most competitive bid and $a\in A$ is the least competitive ask such that $b$ and $a$ are tradable, we need to show the existence of a matching $M_{ab}$ such that $\vol(M_{ab}) = \vol(M)$ and $M_{ab}$ contains a transaction between $b$ and $a$ with transaction quantity $\min\{\q(b),\q(a)\}$. Let $q = \min\{\q(b),\q(a)\}$.

We do the following surgery on $M$ in three steps to obtain the desired $M_{ab}$. 

Step 1: We first modify $M$ to ensure that bid $b$ has at least $q$ total traded quantity in $M$. This is accomplished by running the $\FOB$ procedure on $M$ that outputs a matching that prefers the most competitive bid, which is $b$ in our case, over any other bids. Since $\vol(M) \ge q$, we get that $\FOB(M,B)$ has at least $q$ trade quantity of $b$. Note that $\FOB$ does not change the volume of $M$. Set $M \leftarrow \FOB(M,B)$. 

Step 2: In this step, we modify $M$ to ensure that the bid $b$ and ask $a$ have $q$ quantity traded (not necessarily between them). If $\Q(M,\id(a)) < q$, then there exists a transaction $m$ in $M$ such that bid of $m$ is $b$ and ask of $m$ is not $a$ (since $\Q(M,\id(b))\geq q$). We modify $M$ such that we reduce a single unit from $m$ and increase the trade between $b$ and $a$ by a single unit in $M$. Now the modified matching $M$ has an extra quantity of $a$ without affecting the Volume of $M$ and the total traded quantity of $b$ in $M$. We repeat this process until $\Q(M,\id(a)) = q$.

Step 3: In this step, we modify $M$ to ensure that $b$ and $a$ have a transaction with transaction quantity $q$. Since $Q(M,\id(b))\geq q$ and $Q(M,\id(a))\geq q$, if the the total traded quantity between $b$ and $a$ is strictly less than $q$, then there are two transactions $m_1$ between $b$ and $a'$ ($a'\neq a$) and $m_2$ between $b'$ ($b'\neq b$) and $a$ in $M$. 
We reduce the transaction quantities of $m_1$ and $m_2$ by a single unit each. Next, we increase the transaction quantity of the transaction between $b$ and $a$ by a single unit. Finally, we increase the transaction quantity of the transaction between $b'$ and $a'$ by a single unit; note $b'$ and $a'$ must be tradable, as  $b'$ was traded with $a$, the least competitive ask, so $b'$ is tradable with all asks. Note that this step does not alter $\vol(M)$, $\Q(M, \id(b))$, and $\Q(M,\id(a)$, but increases the transaction quantity between $b$ and $a$. We repeatedly apply this surgery to obtain the desired matching.

\end{proof}

\subsection{Optimal-Uniform Matching Algorithm}
\label{sec:uniform}
In this section, we describe the $\UM$ process that takes as input a list of bids and a list of asks and produces a fair and optimal uniform matching that can be directly used in the financial markets for conducting call auctions. We present a proof of correctness of the $\UM$ process.

\begin{algorithm}[H]
\caption{The $\mathsf{UM}$ Algorithm }\label{algo:um}
\begin{algorithmic} 
\Function{$\UM$}{Bids $B$, Asks $A$}
\State Sort the bids in $B$ in decreasing order of competitiveness.
\State Sort the asks in $A$ in decreasing order of competitiveness.
\State $M = \match(B,A,\emptyset)$
\State $p = \ltp(M)$
\State \Return $\altp(p, M)$
\EndFunction
\end{algorithmic} 
\end{algorithm}

Given the lists of bids and asks, $B$ and $A$, $\UM$ first sorts them (by decreasing competitiveness). 
It then invokes $\match$ on the sorted lists $B$, $A$, and the empty matching $\emptyset$, which outputs a matching $M$. Note that the transaction price assigned by $\match$ to each matched bid-ask pair is the price of the ask in that pair, and hence the matching output by $\match$ need not be uniform.\footnote{Observe that any value in the interval of the limit prices of the matched bid-ask pair can be assigned as the transaction price and it will not affect any analysis done in this work.}
To produce a uniform matching we have to assign a single transaction price to all the transactions of $M$ which we choose to be the transaction price, $p$, of the last matched bid-ask pair, $(b',a')$, of $M$. This is done by the $\mathsf{Assign\_Transaction\_Price}$ subroutine and it does not affect the transaction quantities of $M$; thus, the volume of $M$ is not affected. We now explain why this assignment does not violate the matching property of $M$. Let the participating bids of $M$ be $B'\subseteq B$ and participating asks of $M$ be $A'\subseteq A$. Recall that before assigning the uniform transaction price $M$ is a matching from \cref{lem:match_matching}. 
Since $\match$ scans the sorted lists $B$ and $A$ in decreasing competitiveness, $\pr(a')\geq\pr(a)$ for all $a\in A'$ and $\pr(b')\leq \pr(b)$ for all $b\in B'$. Since, $\pr(a')\leq p \leq \pr(b')$, we have for all $a\in A'$ and $b\in B'$ $\pr(a)\leq\pr(a')\leq p\leq\pr(b')\leq\pr(b)$. Hence, $p$ will not violate the limit prices of any of the matched bid-ask pairs. Thus, $\UM$ outputs a uniform matching.

Next, from \cref{lem:match_foa,lem:match_fob} $\match$ outputs a fair matching since both $B$ and $A$ are sorted. Notice that the fairness property is not affected by updating the transaction prices of a matching. Thus, $M$ remains a fair matching after uniform price assignment. Finally, from \cref{lem:match_optimal_um}, we conclude that $\UM$ outputs a maximum volume uniform matching. Combining all these observations, we obtain our main result for $\UM$.

\UMcorrect*

In our Coq formalization, this theorem appears as follows.

\begin{minted}[bgcolor=gray!05, fontsize=\scriptsize]{coq} 
(* The UM is fair and optimal uniform algorithm. *)
Theorem UM_correct B A:
        admissible B A -> 
        Is_fair (UM B A) B A /\ Is_optimal_uniform (UM B A) B A.
\end{minted}

\subsection{Maximum Matching Algorithm}
\label{sec:maximum}
In this section, we describe the $\MM$ process that takes as input a list of bids $B$ and a list of asks $A$ and produces a maximum volume matching over $(B,A)$. We present a proof of correctness of the $\MM$ process.

\begin{algorithm}[H]
\caption{The $\mathsf{MM}$ Algorithm }\label{algo:mm}
\begin{algorithmic}
\Function{$\MM$}{Bids $B$, Asks $A$}
\State Sort the bids in $B$ in decreasing order of competitiveness.
\State Sort the asks in $A$ in increasing order of competitiveness.
\State $M = \match(B,A,\emptyset)$
\State $M' = \FOA(M,A)$
\State \Return $M'$
\EndFunction
\end{algorithmic} 
\end{algorithm}

 $\MM$ first sorts the list $B$ in decreasing order of competitiveness and list $A$ in increasing order of competitiveness, i.e., the most competitive bid is at the top of $B$ and the least competitive ask is at the top of $A$. It then invokes $\match$ on the lists $B$, $A$, and the empty matching $\emptyset$, which outputs the matching $M$. Finally, it invokes the $\FOA$ algorithm with $M$ and $A$ as the input, to obtain $M'$, which is then returned. We argue that $M'$ is a maximum matching over $(B,A)$.

From \cref{lem:match_matching}, the $\match$ algorithm outputs a matching. From \cref{lem:match_fob} $\match$ outputs a fair on bids matching since $B$ is sorted in decreasing order of competitiveness. Also, from \cref{thm:FOA}, the output $M'$ of $\FOA(M,A)$ is a fair matching. Finally, from \cref{lem:match_optimal_mm}, $M'$ is a maximum volume matching. Thus, our main result for $\MM$ can be stated as follows.

\MMcorrect*

In our Coq formalization, this theorem appears as follows.

\begin{minted}[bgcolor=gray!05, fontsize=\scriptsize]{coq} 
(* The MM is fair and maximum volume matching algorithm. *)
Theorem MM_correct B A:
        admissible B A -> 
        Is_maximum (MM B A) B A /\ Is_fair (MM B A) B A.
\end{minted}

Formalization notes: First we define the `Match' function and prove all of its properties in the `Match.v' file. The $\mathsf{UM}$ process and its correctness proof are written in the `UM.v' file. Similarly, $\mathsf{MM}$ process and its correctness proof are written in the `MM.v' file.

\section{Uniqueness Theorems and Automated Checkers} \label{sec:uniqueness}
In this section, we establish certain theorems that enable us to automatically check for violations in an exchange matching algorithm by comparing its output with the output of our certified program. Detailed proofs are available in the Coq formalization~\cite{git:call}.

Ideally, we would have wanted a theorem that states that the properties (fair and optimal uniform) imply a unique matching. Such a theorem would enable us to automatically compare a matching produced by an exchange with a matching produced by our certified program to find violations of these properties in the matching produced by the exchange. Unfortunately, such a theorem is not possible; there exists two different matchings $M_1$ and $M_2$ over the same admissible order-domain both of which are fair and optimal uniform: $M_1 = \{(b_1,a_1,1,p), (b_2,a_2,2,p)\}$ and $M_2 = \{(b_1,a_2,1,p), (b_2,a_2,1,p), (b_2,a_1,1,p)\}$ on bids $B = \{(b_1,* , 1, p),  (b_2, *, 2, p)\}$ and asks $A = \{(a_1,*,  1, p),  (a_2,*, 2, p)\}$ for some arbitrary price $p$ and timestamps (which are not made explicit; instead we use $*$ as a placeholder). Observe that $M_1$ and $M_2$ are both uniform (since transaction prices are all $p$), fair (since all orders are fully traded in both $M_1$ and $M_2$), with volume $3$ (which is the maximum possible volume as $\vol(B) = 3$). Note that fairness does not require the most competitive bid to be paired with the most competitive ask. For example, assuming $a_1$ has a lower timestamp than $a_2$ and $b_1$ has a lower timestamp than $b_2$ in the above example, $a_1$ and $b_1$ are not matched in the matching $M_2$, which is a fair matching.
Nonetheless,  we can show that given an admissible order-domain, all matchings that are fair and uniform must have the same trade volume for each order. This still allows us to automatically check for violations of the properties in an exchange, by comparing the trades of each order produced by the exchange against that produced by our certified program.

We have the following theorem which formulates this uniqueness relation on the matchings.

\uniquenessTheorem*

This theorem appears in our Coq formalization as 

\begin{minted}[bgcolor=gray!05, fontsize=\scriptsize]{coq} 
(* Uniqueness preperty (completeness) *)
Theorem completeness M1 M2 B A:
        admissible B A /\ (Vol(M1) = Vol(M2)) /\
        (Matching M1 B A) /\ (Matching M2 B A) /\
        Is_fair M1 B A /\ Is_fair M2 B A ->

        (forall a, Qty_ask M1 (id a) = Qty_ask M2 (id a)) /\
        (forall b, Qty_bid M1 (id b) = Qty_bid M2 (id b)).
\end{minted}

Observe that if we specify that an algorithm must output a fair and a maximum volume matching, the output is ``unique'' in the sense that each order in the input will have the same quantity traded in the output matching. Similarly, uniqueness holds when we specify that the algorithm must output a fair and an optimal uniform matching.
The only freedom that the algorithm has is in deciding which bid gets traded with which ask, their trade quantity, and the transaction price. For uniform price matching, which is predominantly used in the opening sessions of various stock markets for price discovery, in fact, who gets traded with whom also becomes practically irrelevant, as all participants are matched at the same price.
Thus, roughly speaking, these requirements form an almost complete specification for the problem.

From the above theorem, the following corollaries are immediate. 
\begin{corollary}\label{thm:uniquenessTheorem2}
For any two fair and maximum matchings $M_1$ and $M_2$  over an admissible order-domain $(B, A)$, for each order $\omega$, the total traded quantity of $\omega$ in $M_1$ is equal to the total traded quantity of $\omega$ in $M_2$.
\end{corollary} 

\begin{corollary}\label{thm:uniquenessTheorem3}
For any two fair and optimal uniform matchings $M_1$ and $M_2$  over an admissible order domain $(B, A)$, for each order $\omega$, the total traded quantity of $\omega$ in $M_1$ is equal to the total traded quantity of $\omega$ in $M_2$.
\end{corollary} 

For optimal uniform matching, for each order, we can compare the total traded quantities of the order in the matching $M_1$ produced by an exchange with the total traded quantities of the order in the matching $M_2=\UM(B, A)$ produced by our certified program. If for some order, the traded quantities do not match, then from \cref{thm:UMcorrect} and  \cref{thm:uniquenessTheorem3} we know that $M_1$ does not have the desired properties as required by the regulators. On the other hand, if they do match for all orders, then the following theorem gives the guarantee that $M_1$ is fair (note that the uniform property can be verified directly from the transaction prices and clearly the total trade volume of $M_1$ and $M_2$ are the same if the traded quantities are the same for each order).

\ConverseUniquenessTheorem*

The above theorem appears in our Coq formalization as follows. 

\begin{minted}[bgcolor=gray!05, fontsize=\scriptsize]{coq} 
(* Converse uniqueness preperty *)
Theorem soundness M1 M2 B A:
        admissible B A /\ 
        (Matching M1 B A) /\ (Matching M2 B A) /\
        Is_fair M2 B A /\ (Vol(M1) = Vol(M2)) /\
        (forall a, Qty_ask M1 (id a) = Qty_ask M2 (id a)) /\
        (forall b, Qty_bid M1 (id b) = Qty_bid M2 (id b)) -> 

        Is_fair M1 B A.
\end{minted}

We now provide the proofs of these theorems. 

\begin{proof}[Proof of \cref{thm:uniquenessTheorem}]
    
We will prove by contradiction using the following property of a matching \begin{equation} \label{eq7} \vol(M) = \sum_{b \in B}\Q(M, \id(b)). \end{equation}

Let $M_1$ and $M_2$ be fair matchings such that $\vol(M_1) = \vol(M_2)$. Let $b$ be a bid whose total trade quantity in $M_1$ is different (without loss of generality, more) from its total trade quantity in $M_2$. It is easy to show that there exists another bid $b'$ such that its total traded quantity in $M_1$ is less than her total traded quantity in $M_2$, i.e., $\Q(M_2, \id(b')) > \Q(M_1, \id(b'))$ (since the sum of the total traded quantities of all the bids of $B$ in $M_1$ is equal to the sum of the total traded quantities of all the bids of $B$ in $M_2$ from \cref{eq7}.

Now, there can be two cases: (i) $b$ is more competitive than $b'$ or (ii) $b'$ is more competitive than $b$. In the first case, since $\Q(M_1, \id(b)) > \Q(M_2, \id(b))$, it follows that $\Q(M_2, \id(b)) < \Q(M_1, \id(b)) \leq \q(b)$; in particular, in the matching $M_2$, $b$ is not fully traded. But, since $\Q(M_2, \id(b'))$ $ > \Q(M_1, \id(b')) \ge 0$, we have that $b'$ gets traded in $M_2$. This contradicts the fact that $M_2$ is fair on the bids as a less competitive bid $b'$ is being traded in $M_2$, while a more competitive bid $b$ is not fully traded. Similarly, in the second case, we can derive a contradiction to the fact that the matching $M_1$ is fair on the bids.
\end{proof} 

The proof of \cref{thm:ConverseUniquenessTheorem} follows immediately from the definition of fairness.

Formalization notes: All the theorems in this section are formalized in the file `Uniqueness.v' using the above proof ideas.

\subsection{Demonstration: Automatic Detection of Violations}
\label{sec:demonstration}

In this section, we demonstrate the practical applicability of our work. For this, we procured real data from a prominent stock exchange. This data consists of order-book and trade-book of everyday trading for a certain number of days. For our demonstration, we considered trades for the top 100 stocks (as per their market capitalization) of a particular day. For privacy reasons, we conceal the real identity of the traders, stocks and the exchange by masking the stock names (to s1 to s100) and the traders' identities. We also converted the timestamps appropriately into natural numbers (which keeps the time in microseconds, as in the original data).
Furthermore, the original data has multiple requests with the same order id; this is because some traders update or delete an existing order placed by them before the call auction is conducted.
After our preprocessing, we just have the final lists of bids and asks in the order-book that participate in the auction. Furthermore, there are certain market orders, i.e., orders that are ready to be traded at any available price, which effectively means a limit price of zero for an ask and a limit price of infinity for a bid; in the preprocessing, we set these limit prices to zero and the largest OCaml integer, respectively.

We then extracted the verified OCaml programs and ran them on the processed market data. The output trades of the verified code were then compared with the actual trades in the trade-book from the exchange. From the uniqueness theorems in \cref{sec:uniqueness}, we know that if the total trade quantity of each order in these two matchings are equal, then the matching produced by the exchange has the desired properties (whether it is uniform or not, can be checked trivially by looking at the prices in the trade-book). We also know that if they are not equal for some traders, then the matching algorithm of the exchange does not have the requisite desired properties (or there is some error in storing or reporting the order-book or the trade-book accurately).

The processed data and the relevant programs for this demonstration are available at ~\cite{git:call}. The extracted OCaml programs of the functions required for this demonstration are stored in a separate file named `certified.ml'. The input bids, asks, and trades of each stock are in `s.bid', `s.ask', and `s.trade' files, respectively, where `s' is the masked id for that stock. For example, file `s1.bid' contains all the bids for the stock `s1'. 

Our automated checker additionally uses two OCaml scripts: create.ml and compare.ml. The create.ml script feeds inputs (lists of bids and asks) to the UM process, and then prints its output matching $M$. The compare.ml script compares the matching produced by the UM process $M$ with the actual trades $M_\text{EX}$ in the exchange trade-book. If the total trade quantity for all the traders in $M$ matches with that of the total trade quantity in $M_\text{EX}$, then the compare.ml script outputs "Matching does not violate the guidelines". If for some bid (or ask) the total trade quantity of $M$ and $M_\text{EX}$ does not match, then the program outputs "Violation detected!".

Out of the 100 stocks we checked, for three stocks our program outputted "Violation detected!". When we closely examined these stocks, we realized that in all of these stocks, a market ask order (with limit price = 0), was not matched by the exchange in its trading output (and these were the only market ask orders in the entire order-book). On the contrary, market bid orders were matched by them. With further investigation, we observed that corresponding to each of these three violations, in the raw data there was an entry of update request in the order-book with a limit price and timestamp identical to the uniform price and the auction time, respectively. It seems highly unlikely that these three update requests were placed by the traders themselves (to match the microsecond time and also the trade price seems very improbable); we suspect this is an exchange's system generated entry in the order-book. We hope that the exchange is aware of this and doing this consciously. When we delete the market asks in the preprocessing stage, no violations are detected. Even if it is not a violation (but a result of the exchange implementing some unnatural rule that we are not aware of), it is fascinating to see that with the help of verified programs, we can identify such minute and interesting anomalies that can be helpful for regulating and improving the exchange's matching algorithm.

\bibliography{main}
\end{document}